\documentclass{PoS}

\title{Three-Field Potential for Soft-Wall AdS/QCD}

\ShortTitle{Three-Field AdS/QCD}

\author{\speaker{Sean P. Bartz }\\
        University of Minnesota\\
        E-mail: \email{bartz@physics.umn.edu}}

\author{Joseph I. Kapusta\\
        University of Minnesota\\
        E-mail: \email{kapusta@physics.umn.edu}}

\abstract{The AdS/CFT correspondence may offer new and useful insights into the non-perturbative regime of strongly coupled gauge
theories such as Quantum Chromodynamics. Soft-wall AdS/QCD models have reproduced the linear trajectories of meson spectra
by including background dilaton and chiral condensate fields. Efforts to derive these background fields from a scalar potential have
so far been unsuccessful in satisfying the UV boundary conditions set by the AdS/CFT dictionary while reproducing the IR
behavior needed to obtain the correct chiral symmetry breaking and meson spectra.

We present a three-field scalar parametrization that includes the dilaton field and the chiral and glueball condensates. This model is consistent with linear trajectories for the meson spectra and the correct mass-splitting between the vector and
axial-vector mesons. We also present the resulting meson trajectories.}

\FullConference{The 7th International Workshop on Chiral Dynamics,\\
		August 6 -10, 2012\\
		Jefferson Lab, Newport News, Virginia, USA}

\begin{document}

\section{Introduction and Motivation}

Quantum chromodynamics has been well tested for high-energy collisions, where perturbation theory is applicable. However, at hadronic scales, the interaction is non-perturbative, requiring a new theoretical model. The Anti-de Sitter Space/Conformal Field Theory (AdS/CFT) correspondence establishes a connection between $n$-dimensional Super-Yang Mills Theory and a weakly-coupled gravitational theory in $n+1$ dimensions \cite {maldacena}. Phenomenological models inspired by this correspondence are known as AdS/QCD, and have succeeded in capturing some features of QCD. A representative, though not exhaustive, list of examples is included here: \cite{stephanov-katz-son, lyubovitskij, kwee-lebed-pion, karch-katz-son-adsqcd, bartz-pions, Brodsky-original}. 

 Quark confinement in QCD sets a scale that is encoded in a cut-off of the fifth dimension in the AdS theory. Soft-wall models use a dilaton as an effective cut-off to limit the penetration of the meson fields into the bulk. The simplest soft-wall models use a quadratic dilaton to recover the linear Regge trajectories \cite{stephanov-katz-son}, while models that modify the UV behavior of the dilaton more accurately model the ground state masses. We use the meson action from \cite{gherghetta-kelley}.
 
 \begin{equation}
 S_{meson}=\int d^{5}x\sqrt{-g}e^{-\Phi(z)}Tr\left[|DX|^{2}+m_{X}^{2}|X|^{2}+\kappa|X|^4+\frac{1}{4g_{5}^{2}}(F_{L}^{2}+F_{R}^{2})\right]. \label{eq:mesonaction}
\end{equation}
Here, $\Phi$ is the dilaton, $X$ is the scalar field, and $F_{L,R}$ contain the vector and axial vector gauge fields. The $z$-dependent vacuum expectation value of the scalar field encodes the chiral symmetry breaking,
 \begin{equation}
 \langle X\rangle = \frac{\chi(z)}{2} I,
 \end{equation}
where $I$ is the $N_F \times N_F$ identity matrix. The field strength tensors and covariant derivative are defined as
\[
F_{L,R}^{MN}=\partial^{M}A_{L,R}^{N}-\partial^{N}A_{L,R}^{M}-i[A_{L,R}^{M},A_{L,R}^{N}]\]
\[
D^{M}X=\partial^{M}X-iA_{L}^{M}X+iXA_{R}^{M}.\]
Please see \cite{gherghetta-kelley, bartz-pions} for details on the field content, the equations of motion, and the numerical methods for calculation of the meson spectra.

The models of \cite{stephanov-katz-son, gherghetta-kelley} use parametrizations for the background dilaton and chiral fields that are not derived as the solution to any equations of motion. A well-defined action provides a set of background equations from which these fields can be derived. In addition, this action provides access to the thermal properties of the model through perturbation of the geometry \cite{kelley-thermo}. The full action contains the meson action (\ref{eq:mesonaction}) and the gravitational action, each of which has an intrinsic coupling to the dilaton,
\[\mathcal{S} = \int d^5x\sqrt{-g}e^{-\Phi} \mathcal{L}_{meson}+ \int d^5x\sqrt{-g}e^{-2\Phi} \mathcal{L}_{grav}.\]
In the Einstein frame, the action for the background fields reads
\begin{equation}
S_{grav}=\frac{1}{16\pi G_5}\int d^{5}x\sqrt{-g_E}\left(R_E-\frac{1}{2}\partial_\mu\phi\partial^\mu\phi-\frac{1}{2}\partial_\mu\chi\partial^\mu\chi-V(\phi,\chi)\right). \label{EinsteinAction}
\end{equation}
The dilaton is scaled for a canonical action, $\phi=\sqrt{8/3}\Phi$.

One background equation does not depend on the potential,
\begin{equation}
6a\phi''(z)+[\phi'(z)]^{2}(6a^{2}-1)-[\chi'(z)]^{2}+\frac{12a\phi'(z)}{z}=0.  \label{V-indep1}
\end{equation}
Examining (\ref{V-indep1}), and requiring the IR behavior that $\phi=\lambda z^2$ and $\chi = \Gamma  z$, determines $a=1/\sqrt{6}$ \cite{Springer2010}. This determines a relationship between $\lambda$, the parameter that sets the slope of the Regge trajectories, and $\Gamma$, which is related to the mass-splitting between the vector and axial vector mesons \cite{gherghetta-kelley}
\begin{equation}
\Delta m^2 = \lim_{z\rightarrow\infty} \frac{g_5^2\chi^2}{z^2}=g_5^2\Gamma^2=g_5^2 6^{3/2}\lambda.
\end{equation}
 Using the experimental value of $\lambda$  gives a value of $\Delta m^2$ that is $\sim50$ times larger than what is observed experimentally. Because this inconsistency arises from a background field that does not depend upon the choice of potential, this problem is pervasive in such models with two background fields.

\begin{figure}[htb]
\centerline{%
\includegraphics[width=11cm]{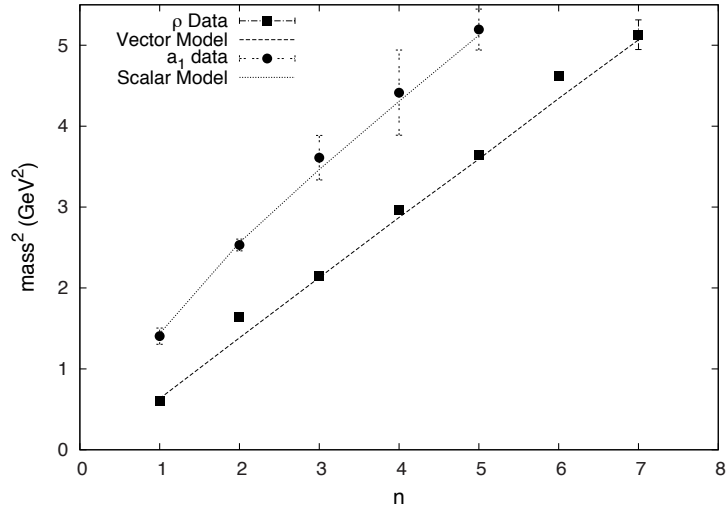}}
\caption{The $\rho$ and $a_1$ meson masses \cite{PDG} are fit well using the three-field parametrization.}
\label{Fig:VA}
\end{figure}
\section{Three-Field Model}

We propose to solve this problem by adding another scalar field to the action for the background fields, $G$,  dual to the glueball field. By this association, the UV boundary condition is $\lim_{z\rightarrow 0} G(z)=g_o z^4$, where $g_o$ is the gluon condensate. To maintain linear confinement, $G\sim z$ in the IR. The background equations that result are
\begin{eqnarray}
\chi'^{2}+G'^{2}&=&\frac{\sqrt{6}}{z^{2}}\frac{d}{dz}\left(z^{2}\phi'\right), \label{background1} \\
\tilde{V}+12&=&\frac{\sqrt{6}}{2}z^{2}\phi''-\frac{3}{2}(z\phi')^{2}-3\sqrt{6}\phi' ,\\
\frac{\partial \tilde{V}}{\partial\phi}&=&3z\phi' ,\label{background3} \\
\frac{\partial \tilde{V}}{\partial\chi}&=&z^{2}\chi''-3z\chi'\left(1+\frac{z\phi'}{\sqrt{6}}\right), \label{background4} \\
\frac{\partial \tilde{V}}{\partial G}&=&z^{2}G''-3zG'\left(1+\frac{z\phi'}{\sqrt{6}}\right), \label{background5}
\end{eqnarray}
where $(')$ indicates differentiation with respect to $z$. Here we have used a conformally transformed potential from the one in (\ref{EinsteinAction}),
$V(\phi,\chi) = e^{2\phi a}\tilde{V}(\phi,\chi).$ The potential depends on $z$ only through the fields, so we may eliminate one of (\ref{background3}-\ref{background5}).

Examining (\ref{background1}) in the IR limit, we see that we can adjust the coefficients  of the dilaton and chiral condensate fields independently. Thus, the addition of a third scalar background field resolves the phenomenological problem present in the two-field models of \cite{Springer2010, batell-gherghetta}.

%\section{Parametrization and Meson Spectra}

\begin{figure}[htb]
\centerline{%
\includegraphics[width=11cm]{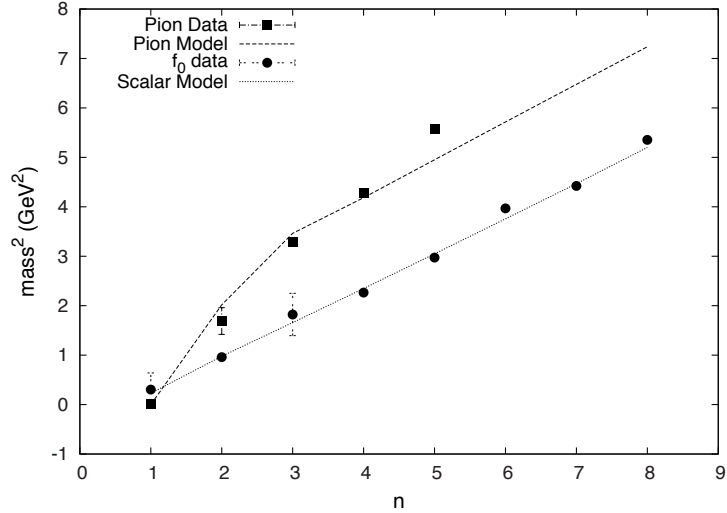}}
\caption{The $f_0$ meson and pion masses \cite{PDG} are fit well using the three-field parametrization. The parameter $\kappa$ is adjusted to avoid a virtual $f_0$ ground state. The pion ground state is massless, due to the zero quark mass in the model. }
\label{Fig:scalar}
\end{figure}

As a check on the three-field setup, we seek a parametrization for the chiral and glueball fields that yields an expression for the dilaton free of special functions. In addition, the parametrization must yield meson spectra that match well with experiment. The following expressions for the derivatives of the chiral and glueball fields match these criteria:
\begin{equation}
G'(z)=\frac{A}{B^3}\left(1-e^{-Bz}\right)^3, \quad \quad
\chi'(z)=\frac{\alpha}{\beta^2}\left(1-e^{-\beta z}\right)^2.
\end{equation}
The parametrization of the chiral field indicates a zero quark mass. The above parameters are defined: $\alpha=\sigma/3, \, \alpha/\beta^2=\Gamma, \, A=g_o/4. $  $B$ is set by ensuring (\ref{background1}) is satisfied in the IR limit.

The values of the parameters, as determined by a least-squares fitting to the $\rho$ and $a_1$ spectra, are: $\sigma = (0.375\, \rm{GeV})^3,\,  g_o=(1.5\,\rm{GeV})^4,\,\Gamma = 0.25\, \rm{GeV}, \, \lambda = (0.428\,\rm{GeV})^2$. The vector and axial-vector meson spectra that result from this parametrization are shown in Figure \ref{Fig:VA}.%, with the scalar and pseudoscalar meson spectra in Figrue \ref{Fig:scalar}.
%, along with the experimental values for the radially excited states of $\rho$ and $a_1$ mesons \cite{PDG}.

The parameter $\kappa$, from the quartic term in the scalar potential in \ref{eq:mesonaction}, is adjusted to give the correct value for the ground state $f_0$ meson. We find that $\kappa =12.4$, in contrast to the negative value found in \cite{gherghetta-kelley}. The scalar potential in the meson action is no longer unbounded from below. There is no virtual ground state for the scalar meson spectrum, which occurs in some models without a quartic interaction term. The results are plotted in Figure \ref{Fig:scalar}.

The numerical results for the pseudoscalar mesons fit the data quite well with the parameters fit to the other sectors. The ground state pion is massless due to the zero quark mass. The linear Regge trajectory is found numerically, without need of a large-$z$ approximation, as in \cite{bartz-pions}. The results are plotted in Figure \ref{Fig:scalar}. 

\subsection{Potentials for Power-Law Background Fields}

We can learn about the asymptotic behavior of the potential by examining its form when the chiral and glueball fields are assumed to have generic power-law behavior:
$ \chi \sim z^n$, $  G \sim z^m$. The dilaton is determined by (\ref{background1}). Adapting the potential ansatz from \cite{Springer2010} to the three-field model,
\begin{equation}
\tilde{V} = -12 + 4\sqrt{6}\phi + c_2\phi^2 -\frac{3}{2}\chi^2+c_3G^2+c_4\chi^4+c_5\phi\chi^2+c_6G^4+c_7\phi G^2+c_8 \chi^2G^2, \label{ansatz}
\end{equation}
and inserting in equations (\ref{background1}-\ref{background5}), we solve for the coefficients $c_i$. There is a unique solution leaving $c_2$ free, which is identified with the dilaton mass as in \cite{Springer2010}.
\begin{eqnarray}
c_3 &=& \frac{-3m-6mn-7m^2n+2m^3n+7mn^2-2mn^3}{2n(1+2m)} \\
c_4 &=& \frac{n^2(c_2-12n(1+n))}{48(1+2n)^2}\\
c_5 &=& \frac{6n^2-c_2n}{2\sqrt{6}(1+2n)}\\
c_6 &=& \frac{m^2(c_2-12m(1+m))}{48(1+2m)^2}\\
c_7 &=& \frac{6m^2-c_2m}{2\sqrt{6}(1+2m)}\\
c_8 &=& \frac{-m(6mn+6n^2+12mn^2-c_2n)}{24(1+2m)(1+2n)}
\end{eqnarray}
Thus, the ansatz (\ref{ansatz}) yields the asymptotic behavior of the potential. However, it remains to find a potential that is consistent with both the UV and IR limits of the background fields. 
%\begin{figure}[htb]
%\centerline{%
%\includegraphics[width=12.5cm]{ScalarFit-scale.pdf}}
%\caption{The experimental $f_0$ meson masses \cite{PDG} are fit well by the eigenvalues found using this parametrization. The parameter $\kappa$ is adjusted to avoid a virtual ground state.}
%\label{Fig:scalar}
%\end{figure}

\section{Conclusion}
We have shown that a gravitational action containing only two background fields does not yield the correct form of chiral symmetry breaking, provided that one of the background fields is identified as the chiral condensate function. In an attempt to circumvent this problem, we have suggested adding a third background field, to be identified with the glueball condensate. A particular parametrization of such a three-field model produces meson spectra that fit the data well. It remains to produce a potential that yields this or another suitable set of background fields.

%\begin{figure}[htb]
%\centerline{%
%\includegraphics[width=12.5cm]{PionFit-scale.pdf}}
%\caption{The experimental pion meson masses \cite{PDG} are fit well by the eigenvalues found using this parametrization. The ground state is massless, due to the zero quark mass in the model.}
%\label{Fig:pions}
%\end{figure}

\bibliographystyle{phys-rev}

\bibliography{seanbartz}

\end{document}